\documentclass[aps,prb,preprint,superscriptaddress,showpacs,endfloats*]{revtex4}
\usepackage[]{graphicx}

\begin{document}

\title{Quantum behavior in nanoscale ballistic rectifiers and artificial
materials}

\author{A. L\"ofgren}
\affiliation{Solid State Physics and Nanometer Consortium,
Lund University, 221 00 Lund, Sweden}
\author{I. Shorubalko}
\affiliation{Solid State Physics and Nanometer Consortium,
Lund University, 221 00 Lund, Sweden}
\author{P. Omling}
\affiliation{Solid State Physics and Nanometer Consortium,
Lund University, 221 00 Lund, Sweden}
\author{A. M. Song}
\thanks{Corresponding author.
Email address: A.Song@umist.ac.uk}
\affiliation{Solid State Physics and Nanometer Consortium,
Lund University, 221 00 Lund, Sweden}
\affiliation{Department of Electrical Engineering and Electronics, UMIST,
Manchester M60 1QD, UK}
\date{\today}

\begin{abstract}
Low-temperature experiments are performed on novel nanoscale nonlinear
devices (ballistic rectifiers) as well as nano-structured artificial
materials, fabricated from an InP/InGaAs quantum well wafer. A
DC output is generated between the lower and upper contacts of these
devices, when an AC
voltage is applied between the left and right contacts. As the
temperature is lowered from room temperature, the DC output voltage of the
ballistic rectifiers gradually changes from negative to positive.  Since the
negative output at high temperatures has been well understood in the
framework of the classical ballistic electron transport, our results
indicate that the electron transport comes into a new physical regime at
low temperatures.  Furthermore, we find that at even lower temperatures, the
devices generate a pronounced oscillatory output as a function of the
applied bias.  Very similar phenomena are observed in the artificial
nanomaterials, suggesting the existence of a common mechanism.  We present a
simple model based on quantum transport, which explains the key phenomena
that we have observed at low temperatures.
\end{abstract}


\maketitle

The rapidly advancing nanotechnology has made it possible to fabricate
devices that are smaller than the electron mean free path, $l_e $.  In
these devices, electrons are mainly scattered at device boundaries rather
than by randomly distributed scatterers such as impurities and phonons, and
it is the device geometry that largely determines the paths traced by the
electrons and, hence, the electronic properties.  Based on this so-called
ballistic electron transport, new device functionalities can be generated
by simply tailoring the device shape.  Although most 
studies on the ballistic electron transport were carried out 
in the linear regime,\cite
{ballistic-reviews} more and more attention is paid to the
non-linear regime.\cite{Lorke-physicaB, AiminPRL98, Linke-europhys,
Linke-sci, Cumming, Novoselov, Hieke, Xu, Worschech} 
Recently, it was also realized that the symmetry of
the device geometry can have a pronounced influence on the nonlinear device
properties. \cite{Lorke-physicaB, AiminPRL98, Linke-europhys,
Linke-sci} 
One such example is the recently realized ballistic rectifier,
in which a triangular antidot was fabricated in a ballistic
micrometer-sized cross junction.\cite{AiminPRL98} The triangular antidot 
breaks the symmetry of the device,
and also serves as an artificial scatterer, which deflects the ballistic
electrons to a certain direction independent of the direction of the
applied electric field.  The generated rectifying functionality has been
shown to be so strong that when ballistic transport was realized at room
temperature by fabricating nanometer-sized ballistic rectifiers, the
sensitivity of microwave detection was found to be roughly as high as a
commercial microwave diode.  \cite{AiminJJAP01} Based on a similar working
principle, artificial functional materials have been made by fabricating an
array of nanometer-sized triangular antidots.  The obtained nanomaterials
have also been demonstrated to operate at room temperature and at
frequencies up to at least 50 GHz.  \cite{AiminAPL01}

Apart from the promising potentials for real applications, the ballistic
rectifiers and the nanomaterials might produce pronounced quantum effects at
low temperatures.  This is because the width of the channels, from which the
ballistic electrons are ejected, is comparable to the electron Fermi
wavelength.  From the physics point of view, studying the influence of
geometric symmetries of nanodevices on the quantum electron transport in the
nonlinear regime is important, and only very limited number of such
experiments have been reported.\cite{Linke-europhys,Linke-sci,Ueno99}
In this work, we report
on temperature-dependent experiments performed on the nanometer-sized
ballistic rectifiers and nanomaterials.  Surprisingly, reversed as well as
oscillatory output is observed when the temperature, $T $, is sufficiently
lowered.  This is in strong contrast to the case at room or high
temperatures, where the experimental results
have been well understood in the framework of
the classical ballistic electron scattering.\cite{Aiminmodel99} The
results indicate that at lower temperatures,
the electron transport in the devices has come
into a new physical regime.  We also discover very similar phenomena in the
nano-structured artificial material.  We propose a simple  model, and
show that the experimental results of the ballistic rectifiers and the
nanomaterials can be qualitatively explained in a
unified physical picture based on the quantum electron transport.

\begin{figure}
\includegraphics[width=7cm]{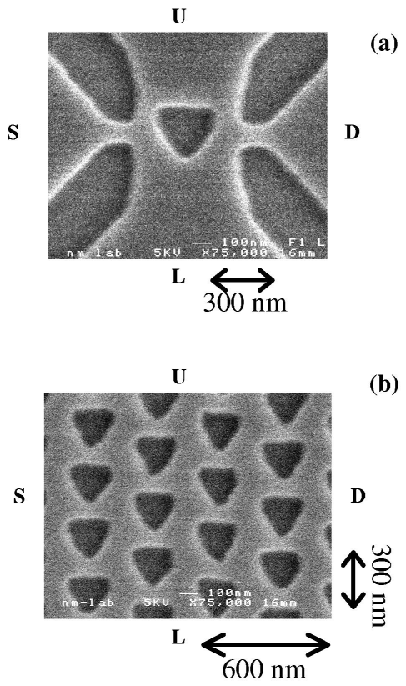}
\caption{Scanning electron micrographs
of (a) the ballistic rectifier and (b) the artificial
nanomaterial measured in the experiments.  The dark triangular areas are
etched down through the layer of the two-dimensional electron gas, and
serve as artificial scatterers for the ballistic electrons.  In both
cases, there are four electrical contacts: left or source (S), right or
drain (D), lower (L), and upper (U).
\label{BR_BRM}
}
\end{figure}

Both the ballistic rectifiers and the nanomaterials are fabricated from
a modulation-doped $\mathrm{In_{0.75}Ga_{0.25}As/InP}$ heterostructure
(details in Ref.~\onlinecite{Ramvall}), in which electrons are confined
to a two dimensional electron gas (2DEG) in a 9 nm thick quantum well
located 40~nm below the surface.  The patterns are defined using
standard electron beam lithography and wet chemical etching (details in
Ref.~\onlinecite{Maximov}).  Figures~\ref{BR_BRM} (a) and (b) show
the scanning electron micrographs of
a typical ballistic rectifier and a part of a typical
nanomaterial measured in the experiments.  The dark triangular areas are
etched down through the 2DEG layer, and serve as artificial scatterers
for the ballistic electrons.  In both cases, there are four electrical
contacts: left or source (S), right or drain (D), lower (L), and upper
(U). 
The width of the source and drain channels of the ballistic rectifier, 
shown in Fig.~\ref{BR_BRM} (a), is about 100~nm, while the width 
of the upper and lower channels is about 500~nm. The triangular antidot
is situated away from the center of the cross junction, which results in 
pronounced nonlinear effects as will be discussed below.  
The nanomaterial, which we show the
results from here, has a repetition period of 600 nm in the horizontal
direction, and 300 nm in the vertical direction.  Both the base and the
height of the triangles are 150~nm.  The device consists of a cross 
junction 
with four contacts, and the written pattern covers the whole area.  The
distance between the contacts is 45~$\mathrm{\mu m}$ both between the
source and the drain, and between the upper and the lower
contacts.
Without illumination, the 2DEG has the following parameters: carrier
concentrations of $4.5 \times 10^{15}$~$\mathrm{m^{-2}}$ and $4.7 \times
10^{15}$~$\mathrm{m^{-2}}$, and mobilities of 45~$\mathrm{m^2/Vs}$ and
1.2~$\mathrm{m^2/Vs}$, at $T=0.3$ K and room temperature, respectively.
Thus, the mean free path $l_e$ varies from 5~$\mathrm{\mu m}$ at 0.3 K
to 140 nm at room temperature.  At room temperature, $l_e$ is shorter
than but comparable to the distance between the triangular antidot
and the source or drain channel in the ballistic rectifier,
meaning that the electron transport is
partially ballistic.  Similarly, the electron transport within
one ``unit cell" of the artificial lattice in Fig.~\ref{BR_BRM}(b),
which largely determines the transport properties of the
whole nanomaterial, is also
partially ballistic at room temperature.  As the temperature is
lowered, the electron transport in both the ballistic rectifier and
one unit cell in the nanomaterial becomes purely ballistic.

\begin{figure}
\includegraphics[width=7cm]{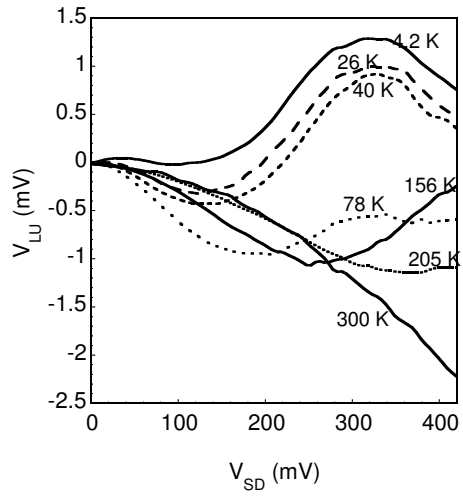}
\caption{The output of the ballistic rectifier,
$V_{LU}$, as functions of the applied source-drain voltage,
$V_{SD}$, measured at different temperatures.
\label{output_BR}
}
\end{figure}

The experiments are performed by applying a DC voltage, $V_{SD}$, between
the source and drain and measuring the DC output voltage, $V_{LU}$,
between the lower and upper contacts.\cite{Effective-output}
Figure~\ref{output_BR} shows the temperature dependence of the ballistic
rectifier.  At room temperature, $V_{LU}$ is
negative, which is consistent with the results of our previous experiments,
\cite{AiminPRL98,AiminJJAP01}, which have been described by a model
based on
the classical ballistic transport.\cite{Aiminmodel99} The output remains
negative as long as the temperature is above about 70 K. However,
starting from about $T= 200$ K, the curves show a clear trend of upward
bending,
and a pronounced output reversal is observed at temperatures below about 40
K. Even more surprisingly, when the temperature is below about 78 K,
oscillatory behavior can be clearly identified.

Separate experiments show that lower temperatures are needed to observe
output reversal if the sample is illuminated by an LED before the
measurements. 
After the illumination, the electron concentration is increased
by up to 30\%
and the electron Fermi wavelength becomes shorter.
A number of devices with different geometric parameters
(from half to about twice the size in Fig.~\ref{BR_BRM}) have also
been measured, and the general trend is that the larger the device, the
lower the temperature at which the output reversal starts to be seen. 
Both facts suggest that quantum effects play an important
role in causing the reversed and oscillatory output. This
may be expected because the width of the source and drain channels,
from which the ballistic electrons are ejected, is comparable to the
electron Fermi wavelength.

Reversed and oscillatory output from a ballistic rectifier
was predicted very recently by Fleischmann and Geisel.\cite{Fleischmann}
The
mechanism was attributed to the large difference between the channel
widths of the upward and downward electron transmissions from S and D
in the ballistic rectifier, i.e.,
the gap between the triangular antidot and the edge of the upper channel is
much narrower than the gap between the triangle and the edge of the lower
channel.  They argued that when the applied voltage is small, the number of
occupied modes  of lateral quantum confinement
in the (lower) wider gap may increase because of the applied
voltage, while the number of
occupied modes in the (upper) narrower gap remains constant due to the
larger energy separation of the lateral confinement energy levels.  This
gives rise to the ``normal'' rectifying effect ($V_{LU}<0$).  When the
applied
voltage is high enough to open up a new mode in the upper gap, however, the
output $V_{LU}$ should undergo a change of sign, i.e., output reversal.

\begin{figure}
\includegraphics[width=7cm]{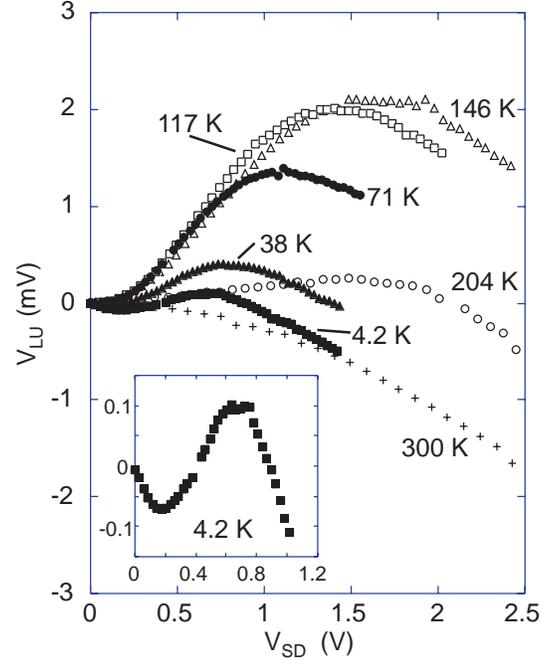}
\caption{ The output of the artificial nanomaterial $V_{LU}$ as functions of
the applied source-drain voltage $V_{SD}$ measured at different
temperatures.  The inset is a magnification of the
4.2~K data.
\label{output}}
\end{figure}

While the result predicted by the model appears to be in good agreement with
the experimental results in Fig.~\ref{output_BR}, such mechanism for the
output reversal does not seem to exist in the nanomaterials shown in
Fig.~\ref{BR_BRM}(b).  This is because that in a nanomaterial, the width of
the channel, which the electrons have to pass to transmit upwards after
being ejected from a narrow gap between two neighboring triangles
in a vertical column, is identical to the channel width of the downward
electron transmission.  As a result, the changes in the occupied lateral
confinement modes in the channels of both upward transmissions and the
downward transmissions are essentially identical.  However, when measuring
the
nanomaterials with various geometric parameters, we observe very similar
reversed as well as oscillatory output to that shown in
Fig.~\ref{output_BR}.  Figure~\ref{output} shows typical results measured
from a nanomaterial, which are unexpected from the model in
Ref.~\onlinecite{Fleischmann}.

In Fig.~\ref{output}, the output is ``normal'' at room temperature, which
has been understood in the framework of the classical ballistic
transport.\cite{AiminAPL01} Output reversal starts to be seen at
temperatures as high as 204 K. A blow-up of the curve at $T=4.2$ K clearly
shows an oscillatory output. \cite{voltage-scale}  Also, similar to the case
of the ballistic rectifiers, the experiments show that the larger the gaps
or the ``lattice constants'' of the artificial lattice, the lower the
temperature at which output reversal starts to be seen.  In
samples with much larger gaps between neighboring triangles (not shown
here), no output reversal is observed down to $T=4.2$ K. This again suggests
that quantum effects play an important role.  Furthermore, the striking
similarities in the experimental results in the ballistic rectifiers and the
nanomaterials suggest the existence of a common physical process that gives
rise to the phenomena.

To understand the output reversal and the oscillatory output, we noticed
that the width of the source and drain channels in the ballistic rectifier
and the width of the gaps between neighboring triangles in a vertical
column in the nanomaterial are all comparable to the electron Fermi
wavelength, which is about 35 nm in our samples.
By taking into account of the
finite depletion depth at the etched device boundaries, the real widths of
the lateral quantum confinement are even narrower than those shown in
Fig.~\ref{BR_BRM}.

\begin{figure}
\includegraphics[width=7cm]{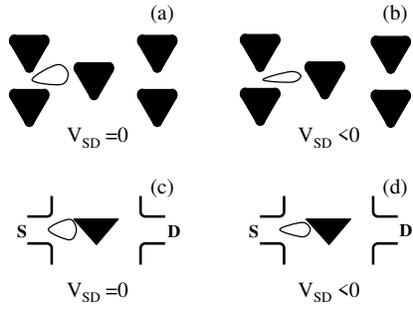}
\caption{
Schematic model of the classical ballistic electron transport at {\em high
temperatures}, for the ``normal'' negative output of the nanomaterial, shown
in (a) and (b), and the ballistic rectifier, shown in (c) and (d).  Note
that the patterns of the angular distribution in the nanomaterial
are tilted a
little bit upwards because the geometry of each gap between the two
neighboring triangles is asymmetric.
\label{high-temp}}
\end{figure}

To simplify the presentation of our model, we first consider the artificial
nanomaterial.  At high temperatures, the transverse quantum confinement
modes in the gaps between neighboring triangles in a vertical column cannot
be well resolved due to the broad Fermi-Dirac distribution.  
The angular distribution of electrons ejected from a gap is not uniform but,
to some extent, collimated,\cite{Beenakkercollimation, Molenkampcollimation,
Shephard92, Hornsey96}, as schematically shown
in Fig.~\ref{high-temp}.  The pattern of the angular distribution
in the nanomaterial is tilted upwards a little because the geometry of
each gap between the two neighboring triangles is
asymmetric.\cite{Hornsey96,Akis96} From the
angular distribution, one anticipates that the triangular scatterer
immediately on the right will deflect most of the electrons downwards, while
fewer electrons can transmit upwards without being scattered by the
triangle.  This may appear to result in an accumulation of electrons in the
lower contact, and consequently induce a negative output voltage $V_{LU}$.
However, this certainly cannot happen if no voltage is applied to the
nanomaterial, since there are always the same number of transmissions along
the opposite directions under the equilibrium condition.  As is shown by the
model of the ballistic rectifier,\cite{Aiminmodel99} a nonzero output can
only be generated by changes of the transmission coefficients caused
by the applied
voltage.  For a finite negative $V_{SD}$, for example, the voltage drops,
mainly at the openings and the exits of the gaps,\cite{Willi}
will cause an increase in the
velocity components of the electrons along the S-D direction, while the
velocity components in the perpendicular direction are virtually not affected.
This leads to smaller ejection angles of the electrons with respect to the
S-D axis, and therefore an overall narrower angular pattern of the electron
flow from S, as shown in Fig.~\ref{high-temp}(b).  Furthermore, the
overall angular distribution of the ejected electrons should also level a
bit as it tends to follow the direction of the electric field.
Such a collimation effect induced by the
applied voltage will clearly change the transmission
coefficients, resulting in more downward deflections of the electrons by
the triangular scatterer, and fewer upward transmissions.
This leads to a ``normal''
negative output $V_{LU}$ as observed at high temperatures in
Fig.~\ref{output} (The probabilities for the electrons from L
and U to transmit to the drain contact are not much affected
because the electric field
is only strong in the region very close to the gaps between neighboring
triangles within a vertical column).
If a positive $V_{SD}$ is applied, the angular pattern
of left-moving electrons from the gaps determines the net flow of carriers,
and should be considered.  Also in this case, the angular pattern
becomes narrower and
more leveled due to the applied electric field, and yields the same negative
$V_{LU}$.  This is also required by the geometric symmetry with
respect to the L-U axis.

At sufficiently low temperatures, the transverse modes in the gaps become
well resolved.  
It is rather complex to determine the number of occupied
modes in the gaps, because while the electron transport between neighboring
triangular antidots is ballistic, the transport over the whole material is
not.  Furthermore, both electron back scatterings and multiple reflections
from the antidots also seriously influence the resistance.  However, from
the width of the gaps (about 100~nm) and the depletion depth (about 10 to 20~nm
at each edge of the etched gaps), 
we estimate that there are between one and four
occupied modes at low temperatures. Separate resistance 
measurements on devices consisting of only a single narrow channel of similar 
width have confirmed the above estimate on the number of occupied modes. 
Similarly to the case of a quantum point contact in the adiabatic transport
regime, each mode contributes to a specific angular pattern of electron flow
from the gaps.  The number of lobes (or branches) in the pattern of electron
flow corresponds to the number of maximums in $|\psi|^2$ (where $\psi$ is
the electron transverse wavefunction), as a direct result of the adiabatic
transport. Therefore, for the Nth transverse
mode, the number of lobes is N. Such branch-like patterns of electron
flow were recently imaged using a low-temperature scanning probe
microscope.\cite{TopinkaScience,Crook00}
In their experiments, additional fine fringes, separated by half 
the Fermi wavelength, were discovered on
the branch-like patterns, which are caused
by coherent constructive and destructive backscattering of the electron 
waves. In general, such effects of quantum phase coherence 
can be easily destroyed by increasing the temperature
or applying a bias voltage
\cite{LinBirdReview02}, which is the reason why the fine fringes 
were observed only at very low effective electron
temperatures. However, the branch-like electron flow survived at 
biases up to at least 3~mV (corresponding to 30--40~K), 
because it is associated with the spacing between the 
transverse quantum confinement subband levels. 
In our experiments, since the 
reversed and oscillatory output is observed at up to about 150~K and 
quite large biases, 
where the phase coherence is unlikely to exist over the distance 
between neighboring triangles, only the branch-like electron flow
is considered and will be shown below 
to result in the observed phenomena.

\begin{figure}
\includegraphics[width=7cm]{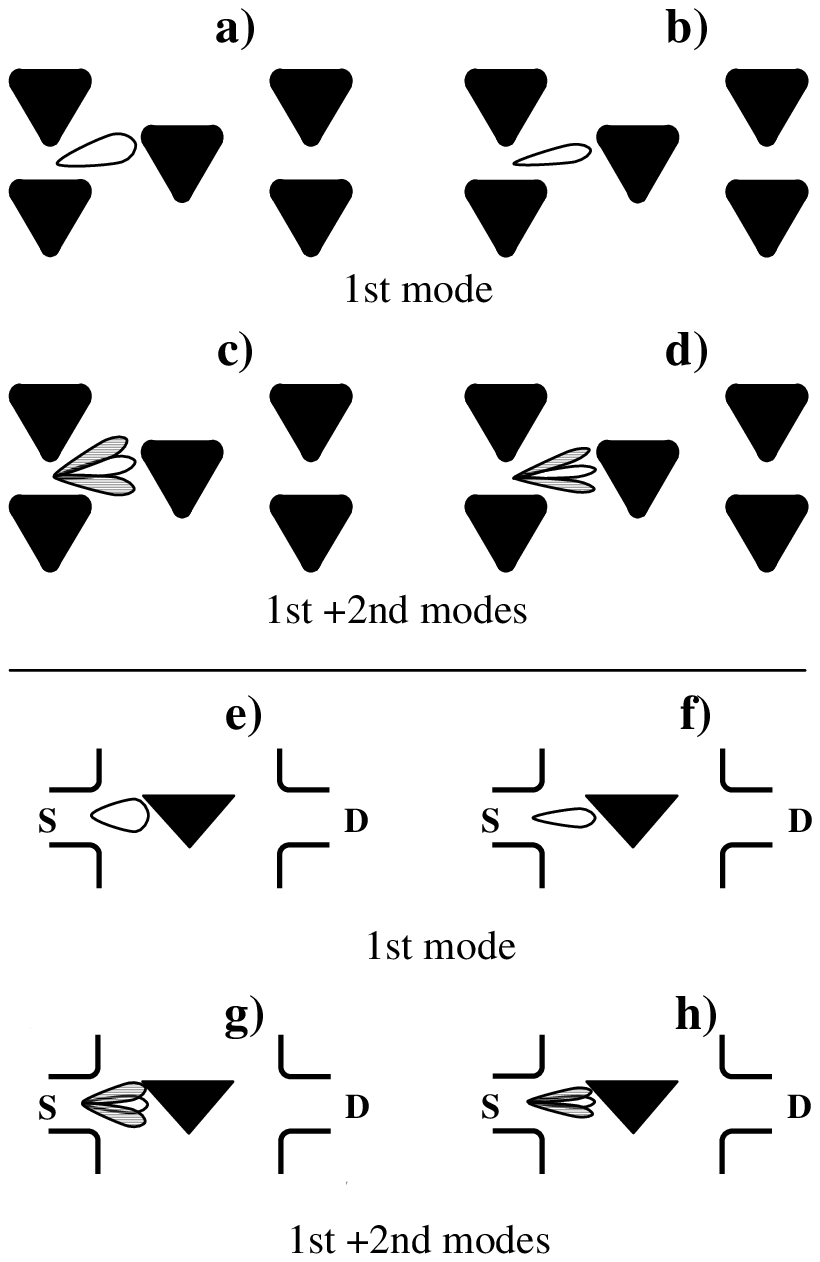}%
\caption{
Schematic model for the output reversal and oscillatory output, in
one unit cell of the artificial nanomaterial (a)-(d) and in the ballistic
rectifier (e)-(h) at {\em low
temperatures}.  As an example, suppose that there is only one lateral
confinement mode occupied in the narrow gaps between neighboring triangles
in a vertical column when the applied voltage $V_{SD}$ is zero.  As a result
of the adiabatic transport at low temperatures, the electrons ejected out of
the gaps have a narrow (rather than uniform) angular distribution (a).  With
decreasing $V_{SD}$, the angular distribution first becomes narrower due to
the collimation effect induced by applied electric field,
and also becomes more
horizontal to follow the direction of the  electric field, as shown
in (b).  By decreasing $V_{SD}$ further, eventually the second lateral
confinement mode becomes occupied, which gives rise to a two-branch angular
distribution, on top of the single-branch pattern of electron flow of the
first occupied mode.  In total, there are three lobes in the electron flow
(c).  The total angular distribution will become narrower again with
continuing decreasing $V_{SD}$ because of the voltage-induced
collimation effect (d). Overall, by decreasing the applied
voltage $V_{SD}$ from zero, the angular distribution of the electrons
ejected out of the gaps changes alternately by subsequentially narrowing
(b), widening(c), narrowing (d)...  Consequently, both the upward and
downward transmission probabilities for the electrons ejected from the gaps
are oscillatory functions of $V_{SD}$, and this induces
the oscillatory output
$V_{LU}$ as shown in Fig.~\ref{output}.  Similar to the case of the
artificial nanomaterial shown in (a)-(d), by decreasing $V_{SD}$ applied to
the ballistic rectifier from zero, the angular distribution of the electrons
ejected out of S changes alternately by subsequentially narrowing (f),
widening(g), narrowing (h),..., because of the interplay between the
voltage-induced collimation and the openings of additional lateral
confinement modes in the source channel.
\label{lobes}}
\end{figure}

For simplicity, we assume that there is only one occupied mode at zero
applied voltage, which corresponds to the angular pattern of the electron
flow shown in Fig.~\ref{lobes}(a), but the following discussion applies to
other (if not too large) numbers of initially occupied modes.  The angular
distribution in Fig.~\ref{lobes}(a) is narrower than that in
Fig.~\ref{high-temp}(a) because of the lower temperature.  We note that
$l_e$ is not longer than a few unit cells of the artificial lattice even at
4.2~K, implying that the electron transport within a unit cell should be, to
a large extent, representative of the overall behavior of the nanomaterial.
If a small negative source-drain voltage is applied, the voltage-induced
collimation effect leads to a narrower and more leveled angular pattern of
the electron flow, which is similar to the case at higher temperatures
shown in
Fig.~\ref{high-temp}(b).  As a result, the downward transmissions of the
electrons deflected by the triangle increase, which gives rise to a negative
output $V_{LU}$.  This explains the initial negative output voltage (or
``normal'' behavior) at low temperatures.  If the applied negative
voltage is
decreased further, the right-moving electrons from the left-hand side of a
gap will eventually occupy the second lateral confinement
mode,\cite{KouwenhovenNonlinear89} which gives
rise to a two-branch angular distribution, on top of the single-branch
pattern of electron flow of the first occupied mode.  In total, there are
three lobes of electron flow as shown in Fig.~\ref{lobes}(c).  The
widening of the angular distribution clearly results in an increased
probability (or percentage)
for the electrons to transmit upwards, and a reduced probability
of the downward transmissions. 
As shown by a detailed model and analysis
in Ref.~\onlinecite{Aiminmodel99} (see Eq.~(6) and the discussions),
the output of a ballistic rectifier
is determined by the {\em relative} changes of the transmission coefficients,
$\Delta T/T$, rather than the absolute values of the changes in
the transmission coefficients $\Delta T$. This should be true for the
artificial material as well,
because of the similarities between a ballistic rectifier
and a unit cell of the artificial material. Therefore, the increased
probability (or percentage) for the electrons to transmit upwards
and the reduced probability of the downward transmissions, which
are discussed above and illustrated in Fig.~\ref{lobes}(c),
will lead to an increase in $V_{LU}$ and
may even cause the output reversal (from negative to positive)
as observed in Fig.~\ref{output} in the
intermediate $V_{SD}$ range. 

If the negative applied voltage is decreased further, the applied voltage or
electric field should again cause the narrowing and leveling of the overall
angular pattern of the electron flow as shown in Fig.~\ref{lobes}(d).  The
right-moving electrons hence have more and more chance to be scattered
downwards by the triangle on the right-hand side, and this contributes again
to a decrease in $V_{LU}$.  The voltage-induced collimation effect is more
pronounced at lower temperatures, since the average kinetic energy of the
electrons is smaller and it is easier to reduce the ejection angle of an
electron at a given voltage drop across the gaps.  At sufficiently low
temperatures the downward bending of the curve may be strong enough to cause
negative output again, and therefore gives rise to an overall oscillatory
output, which is in good agreement with the experimental data in
Fig.~\ref{output}.

The electron transport in the ballistic rectifier is quite similar to that
in the nanomaterial discussed above.  At high temperatures, the collimation
effect induced by a finite negative applied voltage narrows the angular
distribution of the electrons ejected out of S, as shown in
Fig.~\ref{high-temp}(d).  This enhances the probability for the electrons to
be deflected downwards by the triangular antidot, and decreases the
probability of the upward electron transmissions, leading to the ``normal''
negative output $V_{LU}$.  At sufficiently low temperatures, also similar to
the case of the artificial nanomaterial, by decreasing the applied voltage
$V_{SD}$ from zero, the angular distribution of the electrons ejected out of
S changes alternately by subsequentially narrowing as shown in
Fig.~\ref{lobes}(f), widening(g), narrowing (h), due to the interplay
between the electric field induced collimation and the openings of
additional lateral confinement modes in the source channel.
This results in
oscillatory modulations to the transmission probabilities for the electrons
to be transmitted upwards and downwards, and hence leads to an oscillatory
output $V_{LU}$.

Based on this model, it is easy to predict that at low temperatures the
output reversal may well start showing up around $V_{SD}=0$.  In the
above discussion, it is assumed that the Fermi energy at $V_{SD}=0$ lies
well inbetween two transverse subbands, either in the source and drain
channels of the ballistic rectifier or in the gaps between neighboring
triangular antidots in a vertical column in the nanomaterial.  In reality,
it is possible that the Fermi energy is just below the bottom of the next
subband.  Because of the finite broadening of the subband as well as the
Fermi-Dirac distribution, the occupation of the higher subband is not an
abrupt process.  In this case, any finite negative applied voltage may
greatly enhance the population of the high subband, and result in the
broadening of the angular distribution of the ejected electrons and
hence a reversed output starting right from $V_{SD}=0$.  Experimentally,
we indeed observe such a phenomenon, and the low-temperature curves in
Figure~\ref{output_BR} are such an example.  Also, since the chance for the
Fermi energy to be just below a transverse subband is relatively small, in
most cases one should observe a ``normal'' negative output
around $V_{SD}=0$,
which is indeed the case in our experiments where the devices are measured
at different processes of cooling downs, which cause small variations in
the electron concentration and therefore changes in
the electron Fermi energy.

Apart from the similarities in the results of the ballistic rectifier and
the nanomaterials, we notice that there are differences between the
temperature dependences in Figs.~\ref{output_BR} and~\ref{output}.  This is
not surprising because of the noticeable differences between the ballistic
rectifier and the nanomaterial.  So far, we have only discussed the electron
scattering within one unit cell of the nanomaterial.  While a unit cell
should largely represent some of the key transport properties of the whole
nanomaterial, long trajectories of the ballistic electron over more
than one unit cells also play an important role in the determination of the
output voltage $V_{LU}$, especially when the
mean free path is long at low temperatures. The analysis
of the detailed influence of long trajectories of the inter-unit-cell
scatterings is very complex and is not presented in this paper.
Furthermore, as mentioned earlier, the large difference in the channel widths
between the upward and downward transmissions in the ballistic rectifier does
not exist in the nanomaterial.
Therefore, it is expected that the transport properties of the
nanomaterial and the ballistic rectifier can well have
different temperature dependence.

To conclude, we have investigated the electron transport in nanoscale
ballistic rectifiers and nano-structured artificial materials, from the
classical ballistic transport regime at room temperature down to a new
regime at low temperatures. Reversed and oscillatory outputs are
observed.  We present a model based on the quantum and adiabatic electron
transport, which explains the key phenomena that we have observed at low
temperatures, and the similarities in the output from the ballistic
rectifiers and the nanomaterials.

\begin{acknowledgments}
The authors are grateful to L. Samuelson for constant support and
W. Seifert for providing us the InP/InGaAs material.
We wish to thank I. Maximov for the technical help and
H. Q. Xu for the discussions. This work was
supported by the Swedish Research Council, the Swedish Foundation for
Strategic Research, and the European Commission through LTR research
projects Q-SWITCH and NEAR.
\end{acknowledgments}


\begin{thebibliography}{99}
\bibitem{Effective-output} Since the device fabrication cannot be perfect,
the ballistic rectifier and the artificial material are not perfectly
symmetric along the U-L axis.  This causes that the observed output between
the lower and upper contacts, $V_{\rm output}$, at a given applied voltage
$V_{SD}$ is different from the output at $-V_{SD}$.  To eliminate this
effect and focus on the nonlinear effect induced by the artificial triangular
antidot, we always extracted the effective output, ${V}_{LU}(V_{SD}) =
[V_{\rm output}(+V_{SD}) + V_{\rm output}(-V_{SD})]/2$, and reported it
throughout the paper.  This is also equivalent to the case of applying an AC
signal and measure the DC output. Note also that due to the series
resistances
of the ohmic contacts and the broad 2DEG regions between the contacts and
source/drain channels, the real voltage drop on the device is much less
than $V_{SD}$, which also depends on temperature.
\bibitem{voltage-scale} The different voltage ranges of the different curves
shown in Fig.~\ref{output} are due to the fact that the experiments were
carried out by sweeping the source-drain current to a fixed range.  At lower
temperatures, the overall resistances were lower, resulting in smaller
voltage ranges.
\end{thebibliography}
\end{document}